# Continuum analysis of rarefaction effects on a thermally-induced gas flow

Mohamed HSSIKOU [(1)*]; Jamal BALITI[(2)]; Mohammed ALAOUI[(3)]

[(1)] *Faculty of Sciences, Ibn Zohr University, Agadir – Morocco*
[(2)] *Polydisciplinary Faculty, Sultane Moulay Slimane University, Beni Mellal – Morocco*
[(3)] *Faculty of Sciences, Moulay Ismaïl University, Meknes – Morocco*
.
[*]*Corresponding author. E-mail: (hssikoumed@gmail.com)*

A Maxwell gas confined within a micro cavity with non-isothermal walls is investigated in the slip and early transition regimes using the classical and extended continuum theories. The vertical sides of the cavity are kept at the uniform and environmental temperature $T_0$, while the upper and bottom ones are linearly heated in opposite directions from the cold value $T_0$ to the hot one $T_H$. The gas flow is, therefore, induced only by the temperature gradient created along the longitudinal walls. The problem is treated from a macroscopic point of view by solving numerically the so-called regularized 13-moment equations (R13) recently developed as an extension of Grad 13-moments theory to the third order of the Knudsen number powers in the Chapman-Enskog expansion. The gas macroscopic properties obtained by this method are compared with the classical continuum theory results (NSF) using the first and second order of velocity slip and temperature jump boundary conditions. The gas flow behavior is studied as a function of the Knudsen number ($Kn$), nonlinear effects, for different heating rates $T_0/T_H$. The micro cavity aspect ratio effect is also evaluated on the flow fields in this study.

***Keys words:*** R13, NSF, thermal creep, heat flux, rarefaction

## 1. INTRODUCTION:

The frequent use of small-size devices in several applications needs types of equipment that dissipate maximum amount of heat per unit area. With the development of micro-electro-mechanical systems (MEMS), the heat transfer mechanism has gained recently a great interest. The micro pumps, Micro ducts, micro nozzles, micro turbines and micro valves are some typical examples of these small devices involving liquid and gas flows (Graur and Polikarpov 2009, Stone et al. 2004, Yang et al. 2005). Regarding the dimensions of these devices, the continuum assumptions of the gas flow break down and some deviations are observed in respect of the macro scale case. Using velocity slip and temperature jump boundary condition, prove that the rarefaction degree has a significant effect on the gas micro flows or low-pressure gas problems (Karniadakis and Beskok 2005, Marques et al. 2000). To estimate the degree of rarefaction, the so-called Knudsen number defined as $Kn \sim \lambda/L$ is used to classify the flow regimes, where $\lambda$ and $L$ represent respectively the mean free path and a characteristic length of the system. For $Kn < 10^{-3}$ the flow can be considered as a continuum medium, hence the classical Navier-Stokes and Fourier equations are sufficient to describe the gas flow behavior. In the range of $10^{-3} \leq Kn < 0.1$ (slip regime), the gas flow fields can be described using these equations if the typical boundary conditions of velocity slip and temperature jump are applied at the walls (Mizzi et al. 2007). After the slip limit, $0.1 \leq Kn < 10$, i.e., transition regime, the non-equilibrium effects become more important even in the bulk of flow and, therefore, the continuum theory breaks down (Sharipov 2011, Beskok and Karniadakis 1994). In this case, the gas flow must be described from a kinetic point of view by following the governing equation of Boltzmann (2). However, the high-dimensionality of this equation make its direct solving extremely expensive in the discrete phase space. So far, the direct simulation Monte Carlo (DSMC) method is the main practical method used to characterize the strong non-equilibrium flows in NEMS/MEMS (Roohi 2013, Roohi et al. 2009). However, the DSMC method is very expensive both in computational time and memory requirements, especially for low-speed flow in microelectromechanical devices (MEMS) and nanodevices (Fan and Shen 2001). In thermal cavities, when $Kn$ number increases the intermolecular and molecular-surface interactions decrease (Moghadam et al. 2014). So, it is needed huge number of particles to capture the flow features. In the macroscopic point of view, the problems involving Couette and Fourier flows have been treated enough with isothermal walls in the previous study (Sharipov and Seleznev 1998, Sharipov 2002, Garcia and Siewert 2009, Rana et al. 2012). However, minor attempts have been observed for treating the problems with non-isothermal walls boundary conditions (Zhu et al. 2017, Kosuge et al. 2011, Vargas et al. 2014). In this case, a thermal slip gas flow is induced under the effects of *transpiration* or *thermal creep* flows, i.e., the gas is forced into motion at the boundaries. The thermal creep flow of a rarefied gas within an enclosure is treated deterministically by solving, in one hand, the Boltzmann-Shakhov model equation and using DSMC as a stochastic approach in the range of $10^{-2} < Kn < 10$ (Kosuge et al. 2011). At the second one, in such problem, a competition between thermal creep flow and thermal stress flow induced by a gradient of heat flux in the bulk is investigated as a function of rarefaction degree ($Kn$) (Mohammadazadeh et al.2015). At the micro scale and rarefied conditions, the temperature inhomogeneity of a gas system can lead to a variety of flow phenomenon such the *ghost effects* observed at the continuum limit (Sone 2007). The fact that many of these micro-devices involve slip and early transition regime gas flows is the first motivation of this continuum study. The second one is to evaluate the *thermal slip* contribution in the velocity slip and temperature jump phenomena near the heated walls, which is proportional to the tangential temperature-gradient (Sharipov 2016). This effect is the main correction made by the second order of velocity slip and temperature jump in the NSF approach. Investigating thermal behaviors under rarefied condition due to high power densities requires the application of efficient techniques to predict allowable performance limits of these systems (Birur et al. 2001). By comparing the DSMC method and R13 solution, it shown that the R13 equations can be solved in short computation time as compared to those for DSMC method (Rana et al. 2015). Thus, the main goal of this paper is to investigate the rarefaction effects on the macroscopic proprieties of a dilute gas flow induced only by non-isothermal walls effects, thermal creep flow, using the classical and extended continuum-based model. The regularized 13-moment equations (R13), developed by (Struchtrup 2005), are solved numerically to capture the strong rarefaction or nonlinear effects that occur in the early transition regime. In this study, we discuss also the validity of the first (NSF1) and second (NSF2) orders of velocity slip and temperature jump boundary conditions in the Navier-Stokes and Fourier solution, within the rarefaction range of $0.05 \leq Kn \leq 0.3$. In this study, we assume that the gas is not subjected to any external body force. In the next section, we present a brief description of the problem, after we recall the set of basic equations and the boundary conditions used in both solutions of NSF and R13.

## 2. STATEMENT OF PROBLEM:

A Maxwell and monatomic gas is confined within a two-dimensional heated microcavity with orthogonal cross section of $H \times L$ is shown in Fig.1. The left and right walls of the cavity are kept at a uniform and environmental temperature $T_0 = 273K$ while the bottom and upper sides are linearly heated from the cold value $T_0$ to hot one $T_H$ in opposite directions, i.e., $T(x)|_{y=0,H} = T_{0,H} \pm (T_H - T_0)x/L$. The macroscopic properties of the gas are evaluated for different values of Knudsen number in the slip and early transition regimes for the hot temperatures $1.25T_0, 1.5T_0, 1.75T_0$ and $2T_0$. Two values of the cavity aspect ratio, $AR = H/L$, are considered in this study ($AR = 2$ and $1$). The following figure shows the configuration of the cavity in a square geometry ($AR = 1$). The flow fields are evaluated as a function of Knudsen number $Kn$ defined as:

$$Kn = \frac{\mu_0 \, v_0}{p_0} \frac{1}{L}$$

Where $R$, $\mu_0$ and $p_0$ denote, respectively, the gas specific constant, the reference viscosity at the temperature $T_0$ and the hydrostatic pressure given by $p_0 = \rho_0 R T_0$. Also $v_0 = \sqrt{2RT_0}$ is the most probable molecular velocity which is taken as the characteristic velocity.

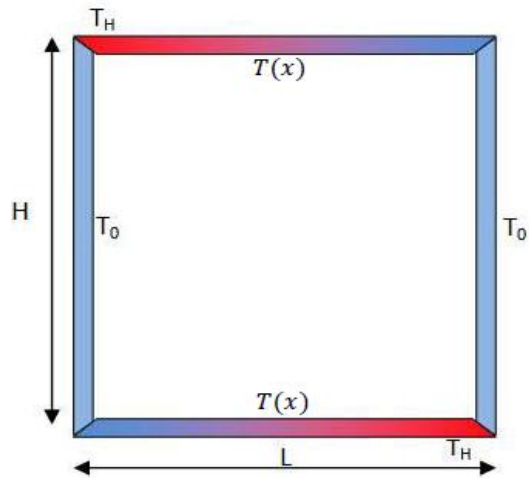

Fig. 1: Micro cavity geometry

## 3. CONTINUUM MODEL BASED DESCRIPTION:

The evolution equations for density, velocity and internal energy ε are given by the conservation laws of mass, momentum and energy that can be written in the general form as follows:

$$\frac{\partial \rho}{\partial t} + \frac{\partial \rho v_i}{\partial x_i} = 0,$$

$$\frac{\partial \rho v_i}{\partial t} + \frac{\partial (P_{ik} + \rho v_i v_k)}{\partial x_k} = 0, \qquad (1a,b,c)$$

$$\frac{\partial \rho \left(\varepsilon + \frac{1}{2}v^2\right)}{\partial t} + \frac{\partial \left(\rho(\varepsilon + \frac{1}{2}v^2)v_k + P_{ik}v_i + q_k\right)}{\partial x_k} = 0,$$

where $\rho, v_k, P_{ik}$ and $q_k$ denote respectively mass density, macroscopic velocity, pressure tensor, and heat flux vector components. Note that the pressure tensor is related to the hydrostatic pressure and stress tensor by the relation: $P_{ij} = p\delta_{ij} + \sigma_{ij}$ where $\delta_{ij}$ represents the Kronecker delta. In the frame of ideal gas approximation, the hydrostatic pressure is given by $p = \rho RT = \rho\theta$ where $\theta$ corresponds to the temperature in energy units. To close this set of equations, one needs to define the constitutive equations including the relationship between the stress tensor **σ**, the heat flux vector **q** and the macroscopic variables derivatives. As an extension of the classical NSF theory, Grad introduces the balance equations corresponding to higher order moments of the particle velocity distribution function $f(\mathbf{r}, \mathbf{c}, t)$ (Grad 1949) which the evolution is governed by the Boltzmann equation given by:

$$(\frac{\partial}{\partial t} + \mathbf{c}.\boldsymbol{\nabla}_\mathrm{r})f = J[f, f'], \qquad (2)$$

where $J[f, f']$ and **c** are respectively the collisions operator and the molecular velocity.

Due to the collision's random nature, the exact analytic solution of the equation (2) remains difficult. For this reason, some approximations are used to simplify this term for the small deviations in respect of the equilibrium state. In this context, it has been shown recently that the Grad's approach accuracy can be further improved up to $\mathcal{O}(Kn^3)$ of the Knudsen number powers in the Chapman-Enskog expansion by regularization of the moment equations of Grad (Struchtrup and Torrilhon 2003, Mizzi et al. 2008). Note that, the NSF solution with velocity slip and temperature jump boundary conditions leads to the first order in this expansion.

In the 13-moments theory of Grad, the gas state is described by the set of 13 moments of functions $\varphi_A = m\{1, c_i, \frac{1}{2}C^2, C_{\langle i}C_{j\rangle}, \frac{1}{2}C^2C_i\}$. The Multiplication of the equation (2), at the steady state, by $\varphi_A$ and subsequent by integration over velocity space yields to the governing equations of moments: mass density, hydrodynamic (bulk) velocity, hydrostatic pressure, stress tensor, temperature and heat flux vector related to the distribution function by:

$$\rho(x,y) = m \int f(\mathbf{r},\mathbf{c}) \, d\mathbf{c}, \qquad (3)$$

$$\mathbf{v}(x,y) = \frac{m}{\rho(x,y)} \int \mathbf{c} f(\mathbf{r},\mathbf{c}) \, d\mathbf{c}, \qquad (4)$$

$$p(x,y) = \frac{m}{3} \int C^2 f(\mathbf{r},\mathbf{c}) \, d\mathbf{c}, \qquad (5)$$

$$P_{ij}(x,y) = m \int C_{\langle i} C_{j \rangle} f(\mathbf{r},\mathbf{c}) \, d\mathbf{c}, \qquad (6)$$

$$T(x,y) = \frac{m}{3nK_B} \int C^2 f(\mathbf{r},\mathbf{c}) \, d\mathbf{c}, \qquad (7)$$

$$\mathbf{q}(x,y) = \frac{m}{2} \int C^2 \mathbf{C} f(\mathbf{r},\mathbf{c}) d\mathbf{c}. \qquad (8)$$

where $\mathbf{C} = \mathbf{c} - \mathbf{v}$ is the peculiar velocity.

By setting $\varphi_A = m$, $mc_i$ and $\frac{1}{2}mC^2$ respectively and using the invariance property of the collisions operator, the right side of (2), one can find the balance equations (1$a,b,c$). The two other values of $\varphi_A$ lead to the nonlinear constitutive equations of stress tensor $\sigma_{ij}$ and heat flux vector $q_i$, respectively that given in the so-called R13 approach by (Struchtrup and Torrilhon 2003):

$$\frac{\partial \sigma_{ij}}{\partial t} + \frac{\partial \sigma_{ij} v_k}{\partial x_k} + \frac{4}{5}\frac{\partial q_{\langle i}}{\partial x_{j \rangle}} + 2p \frac{\partial v_{\langle i}}{\partial x_{j \rangle}} + 2\sigma_{k \langle i} \frac{\partial v_{j \rangle}}{\partial x_k} + \frac{\partial m_{ijk}}{\partial x_k} = -\frac{p}{\mu}\sigma_{ij}, \quad (9)$$

$$\frac{\partial q_i}{\partial t} + \frac{\partial q_i v_k}{\partial x_k} + \frac{5}{2} p \frac{\partial \theta}{\partial x_i} + \frac{5}{2}\sigma_{ik}\frac{\partial \theta}{\partial x_k} + \theta \frac{\partial \sigma_{ik}}{\partial x_k} - \sigma_{ik}\frac{\theta}{\rho}\frac{\partial \rho}{\partial x_k} - \frac{\sigma_{ik}}{\rho}\frac{\partial \sigma_{kl}}{\partial x_l} + \frac{7}{5} q_k \frac{\partial v_i}{\partial x_k} + \frac{2}{5} q_k \frac{\partial v_k}{\partial x_i} + \frac{2}{5} q_i \frac{\partial v_k}{\partial x_k}$$

$$+ \frac{1}{2}\frac{\partial R_{ik}}{\partial x_k} + \frac{1}{6}\frac{\partial \Delta}{\partial x_i} + m_{ikl} \frac{\partial v_k}{\partial x_l} = -\frac{2}{3}\frac{p}{\mu} q_i, \qquad (10)$$

The indices in the angular brackets denote the symmetric trace-free parts of tensors. The R13 equations closure contains, therefore, additional quantities $m_{ijk}$, $R_{ik}$ and $\Delta$, corresponding to the higher order moments, given by (Struchtrup 2005, Mizzi et al. 2008, Rana et al. 2013):

$$\Delta = -\frac{\sigma_{kl}\sigma_{kl}}{\rho} + 6\frac{\sigma_{kl}\sigma_{kl}^{NSF}}{\rho} + \frac{56}{6}\frac{q_k q_k^{NSF}}{p} - 12\frac{\mu}{p}\left(\theta \frac{\partial q_k}{\partial x_k} - \theta q_k \frac{\partial \ln p}{\partial x_k}\right),$$

$$R_{ij} = -\frac{4}{7}\frac{\sigma_{k \langle i}\sigma_{j \rangle k}}{\rho} + \frac{24}{7}\frac{\sigma_{k \langle i}\sigma_{j \rangle k}^{NSF}}{\rho} + \frac{192}{75}\frac{q_{\langle i} q_{j \rangle}^{NSF}}{p} - \frac{24}{5}\frac{\mu}{p}\left(\theta \frac{\partial q_{\langle i}}{\partial x_{j \rangle}} - \theta q_{\langle i} \frac{\partial \ln p}{\partial x_{j \rangle}}\right), \quad (11a-c)$$

$$m_{ijk} = \frac{8}{15}\frac{\sigma_{\langle ij} q_{k \rangle}^{NSF}}{p} + \frac{4}{5}\frac{q_{\langle i} \sigma_{jk \rangle}^{NSF}}{p} - 2\frac{\mu}{p}\left(\theta \frac{\partial \sigma_{\langle ij}}{\partial x_{k \rangle}} - \theta \sigma_{\langle ij} \frac{\partial \ln p}{\partial x_{k \rangle}}\right).$$

If one neglects these quantities, the R13 set of equations will be reduced to the well-known set of Grad's 13-moment equations. At the small values of $Kn$ ($Kn \ll 1$), the stress tensors and heat flux components can be written in the chapman-Enskog expansion of Knudsen number powers as:

$$\sigma_{ij} = \sigma_{ij}^{(0)} + Kn \sigma_{ij}^{(1)} + Kn^2 \sigma_{ij}^{(2)} \ldots \qquad (12)$$

$$q_i = q_i^{(0)} + Kn q_i^{(1)} + Kn^2 q_i^{(2)} \ldots \qquad (13)$$

The zeroth order of this expansion leads to the Euler equation while the first one corresponds to the classical NSF theory. In this case, the Newton and Fourier laws respectively give the stress and heat flux vector for Maxwell gas:

$$\sigma_{ij}^{NFS} = -2\mu \frac{\partial v_{\langle i}}{\partial x_{j\rangle}}, \quad q_i^{NFS} = -\frac{15}{4}\mu \frac{\partial \theta}{\partial x_i}. \quad (14a,b)$$

As in any problem involving partial differential equations, the solving of R13-moment equations requires a set of boundary conditions that one must specify. In this context, Torrilhon and Struchtrup have shown kinetically that $v_n$, $\sigma_{\tau n}$, $q_n$, $R_{\tau n}$, $m_{nnn}$ and $m_{\tau\tau n}$ are the only components that can be prescribed (Torrilhon and Struchtrup 2008a). The subscripts $\tau$ and $n$ denote the tangential and normal components of the tensors respectively, where the walls normal is pointing toward the gas. Using the Maxwell accommodation model for boundary conditions results in a kinetical link between the moments in front of the wall and the tangential wall velocity $v_\tau^w$ and the wall temperature $\theta^w$, the R13-BCs are written as:

$$v_n = 0$$

$$\sigma_{\tau n} = -\chi/(2-\chi)\sqrt{\frac{2}{\pi\theta}}\left(\mathcal{P}\mathcal{V}_\tau + \frac{1}{5}q_\tau + \frac{1}{2}m_{\tau nn}\right),$$

$$q_n = -\chi/(2-\chi)\sqrt{\frac{2}{\pi\theta}}\left(2\mathcal{PT} - \frac{1}{2}\mathcal{PV}_\tau^2 + \frac{1}{2}\theta\sigma_{nn} + \frac{1}{15}\Delta + \frac{5}{28}R_{nn}\right),$$

$$R_{\tau n} = \chi/(2-\chi)\sqrt{\frac{2}{\pi\theta}}\left(6\mathcal{PTV}_\tau + \mathcal{P}\theta\mathcal{V}_\tau - \mathcal{PV}_\tau^3 - \frac{11}{5}\theta q_\tau - \frac{1}{2}\theta m_{\tau nn}\right), \quad (15a-f)$$

$$m_{nnn} = \chi/(2-\chi)\sqrt{\frac{2}{\pi\theta}}\left(\frac{2}{5}\mathcal{PT} - \frac{3}{5}\mathcal{PV}_\tau^2 + \frac{7}{5}\theta\sigma_{nn} + \frac{1}{75}\Delta - \frac{1}{14}R_{nn}\right),$$

$$m_{\tau\tau n} = -\chi/(2-\chi)\sqrt{\frac{2}{\pi\theta}}\left(\frac{1}{5}\mathcal{PT} - \frac{4}{5}\mathcal{PV}_\tau^2 + \frac{1}{14}R_{\tau\tau} + \theta\sigma_{\tau\tau} - \frac{1}{5}\theta\sigma_{nn} + \frac{1}{150}\Delta\right).$$

where $\mathcal{P} = \rho\theta + \frac{1}{2}\sigma_{\tau\tau} - \frac{1}{120}\frac{\Delta}{\theta} - \frac{1}{28}\frac{R_{\tau\tau}}{\theta}$.

Note that $\mathcal{V}_\tau = v_\tau - v_\tau^w$ and $\mathcal{T} = \theta - \theta^w$ represent respectively the velocity slip and temperature jump at the vicinity of the walls. The parameter $\chi$ is the Maxwell accommodation coefficient which its value is $\chi = 1$ for the full diffuse reflection and $\chi = 0$ for the specular one. The effect of such coefficient on thermal transpiration phenomena is evaluated using velocity dependent Maxwell (VDM) boundary condition, and by means of DSMC method (Mohammadzadeh and Struchtrup 2015). A heated microcavity with specular walls is also treated by means of NSF and moments theories (Baliti et al. 2017, 2016).

## 4. NUMERICAL SCHEME

To solve the above differential equations $(1, 9-11)$ in the steady state, it is more convenient to rewrite the equations in the condensed and matrix form:

$$A(\omega)\frac{\partial \omega}{\partial x} + B(\omega)\frac{\partial \omega}{\partial y} + \frac{1}{Kn}P(\omega)\omega = 0. \quad (16)$$

where $\omega = [\rho, v_x, v_y, \theta, q_x, q_y, \sigma_{xx}, \sigma_{xy}, \sigma_{yy}, R_{xx}, R_{xy}, R_{yy}, m_{xxx}, m_{xxy}, m_{xyy}, m_{yyy}, \Delta]$ is the vector of field variables; $A(\omega)$, $B(\omega)$ and $P(\omega)$ are respectively the coefficient matrices in $x, y-$ directions and the production matrix. The matrix equation (16) is solved by means of finite difference approach. The cavity domain is discretized on a network of nodes in both directions. The R13 solution is obtained using the boundary conditions prescribed by a set of equations $(15a - f)$. In the slip flow regime, the Navier-Stokes and Fourier equations solution must be obtained with slip and jump boundary conditions. By ignoring the higher order terms $m_{ijk}$, $R_{ik}$, $\Delta$ and replacing the stress tensor and heat flux vector by their corresponding expressions of NSF, as indicated in the equations ($14a, b$). Thus, the first order of NSF-BCs can be written, therefore, as follows:

$$v_n = 0,$$
$$\sigma_{\tau n}^{NSF(1)} = -\chi/(2-\chi)\sqrt{\frac{2}{\pi\theta}}\left(\mathcal{P}\mathcal{V}_\tau - \frac{3}{4}\mu\frac{\partial\theta}{\partial x_\tau}\right), \qquad (17a-c)$$
$$q_n^{NSF(1)} = -\chi/(2-\chi)\sqrt{\frac{2}{\pi\theta}}\left(2\mathcal{P}\mathcal{T} - \frac{1}{2}\mathcal{P}\mathcal{V}_\tau^2 - \mu\theta\frac{\partial v_{\langle n}}{\partial x_{n\rangle}}\right),$$

where $\mathcal{P} = \rho\theta + \frac{1}{2}\sigma_{\tau\tau}$.

While, the second order of these BCs can be obtained using the second order of stress tensor and heat flux vector described by the Burnett equations (Torrilhon and Struchtrup 2004, Struchtrup 2005):

$$v_n = 0,$$
$$\sigma_{\tau n}^{NSF(2)} = -\chi/(2-\chi)\sqrt{\frac{2}{\pi\theta}}\left(\mathcal{P}\mathcal{V}_\tau + \tfrac{1}{5}q_\tau^{NSF} + \frac{m_{\tau nn}^{(2)}}{2} + \frac{1}{5}\frac{\mu^2}{\rho}\left(\frac{45}{16}\frac{\partial^2 v_\tau}{\partial x_k \partial x_k} - \frac{13}{4}\frac{\partial^2 v_k}{\partial x_k \partial x_\tau}\right)\right) - \frac{1}{18}\frac{R_{\tau n}^{(2)}}{\theta},$$
$$q_n^{NSF(2)} = -\chi/(2-\chi)\sqrt{\frac{2}{\pi\theta}}\left(2\mathcal{P}\mathcal{T} - \frac{1}{2}\mathcal{P}\mathcal{V}_\tau^2 + \tfrac{1}{2}\theta\sigma_{nn}^{NSF} + \frac{\Delta^{(2)}}{15} + \frac{13 R_{nn}^{(2)}}{63}\right) \qquad (18a-c)$$
$$- \frac{\mu^2}{\rho}\left(\frac{45}{16}\frac{\partial^2 v_n}{\partial x_k \partial x_k} - \frac{13}{4}\frac{\partial^2 v_k}{\partial x_k \partial x_n}\right),$$

The terms with the superscript $^{(2)}$ refer to the second-order corrections which are obtained from the R13 constitutive relations (11) by replacing stress tensor and heat flux vector by their NSF expressions; these terms are given as follows:

$$R_{ij}^{(2)} = -\frac{24}{5}\frac{\mu}{\rho}\frac{\partial q_{\langle i}^{NSF}}{\partial x_{j\rangle}}, \quad m_{ijk}^{(2)} = -2\frac{\mu}{\rho}\frac{\partial \sigma_{\langle ij}^{NSF}}{\partial x_{k\rangle}}, \quad \Delta^{(2)} = -12\frac{\mu}{\rho}\frac{\partial q_k^{NSF}}{\partial x_k}. \qquad (19a-c)$$

For more convenience, the results are shown with the dimensionless variables using the following normalizations:

$x, y = x, y/L$, $V_i = v_i/v_0$, $q_i = q_i/(\rho_0\theta_0 v_0)$, $\bar{\rho} = \rho/\rho_0$, $\bar{\theta} = \theta/\theta_0 \equiv T/T_0$, $\bar{\sigma}_{ij} = \sigma_{ij}/(\rho_0\theta_0)$, $\bar{R}_{ij} = R_{ij}/(\rho_0\theta_0^2)$, $\bar{\Delta} = \Delta/(\rho_0\theta_0^2)$ and $\bar{m}_{ijk} = m_{ijk}/(\rho_0\theta_0 v_0)$.

## 5. RESULTS AND DISCUSSION:

The behavior of gas flows in the slip and early transition regimes, often encountered in MEMS devices, is one of the challenging problems which need further investigations. In this regime, the mean free path of gas is at the same order of or even much larger than the system characteristic length, the continuum assumption will break down, and the traditional CFD techniques will, therefore, lead to large errors (Karniadakis et al. 2005). The rarefaction effects influence the flow, such as heat flux induced by gradients of stresses even if there is no temperature gradient (Kogan et al. 1976). For validation purpose, firstly, we compare the DSMC and R13 results by considering the test case treated by Vargas et al. (Vargas et al. 2014). In this case, the gas flow is induced, in one hand, by the thermal creep force created along the non-isothermal walls and by a temperature gradient between the other walls. To stay in the considerations of macroscopic approaches, a small temperature gradient and moderates Knudsen number values are considered in this test case, $T_0/T_H = 0.9$, $Kn = 0.01$ and $0.1$. In this case, the gas flow is subjected to two thermal forces. A *transpiration* force due to the linear profile imposed to the walls-temperature and *thermal stress* induced by the temperature gradient in bulk of flow. These forces may be acting the gas flow in the same or opposite directions. Fig. 2 shows the streamlines overlaid on the temperature contours obtained by both approaches.

R13 and DSMC predict two similar primary vortices in the flow field which push the flow from cold to hot region at the side walls. Kinetically, these patterns are interpreted by the fast motion of hot particles compared to particles coming from the colder region, which are slower. As a reaction, the wall is pushed towards the colder region, or, when the wall is at rest, the rarefied gas is driven from cold to hot, which is observed here. Both approaches give almost similar temperature distribution in the cavity.

In contrast with the classical problems of viscous flows induced by the relative motion of walls, this study focuses on the pure thermal creep flow induced only, in opposite directions, by the effects of non-isothermal walls as shown in Fig. 1. To stay in the frame of small deviations from equilibrium state, we restrict our attention to the macroscopic study of gas flow fields for a moderate temperature $T_H = 1.5T_0$ (Papadopoulos and Rosne 1995). Then, we evaluate the effect of $T_H$ on the maximum of tangential thermal-velocity along the bottom walls. By analogy with the viscous Couette flow (Hssikou et al. 2016a, 2016b), the present problem of "thermal Couette flow" is treated by means of classical approach of NSF, with first and second orders of slip and jump BCs, and extended continuum theory.

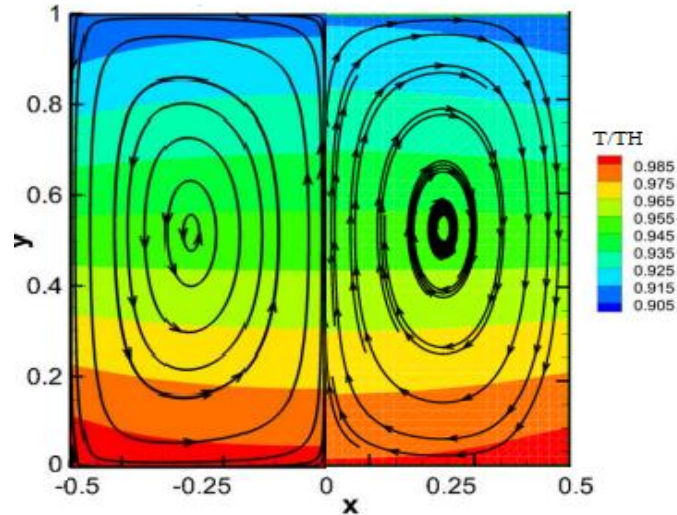

Fig. 2: Streamlines and temperature contours comparison for $T_0/T_H = 0.9$ and $Kn = 0.1$ between R13 (right) and DSMC method [Vargas et al. 2014] (left) adapted to the Vargas's problem (vertical, top-to-down, linearly heated cavity).

The breaks down of the classical theory between the slip and transition regimes limit, $Kn = 0.1$, is clarified by solving the regularized 13-moment equations ($R13$). At low rarefaction degree, $Kn = 0.05$, the $R13$ approach predicts a primary vortex centered at $x = y = 0.5L$ while the $NSF(1)$ and $NSF(2)$ solutions envisage, inside this main vortex, two symmetric secondary vortices corresponding to the flow stagnation. Near longitudinal sides, the tangential temperature-gradient forces a creep-driven flow in a circulatory motion from the cold to hot region. In addition to these eddies induced by the thermal convection phenomenon, thermal creep flow, two secondary symmetric counter-circulating appear at the hot cavity corners are observed, see Fig. 3(a, c, and e). The inverted transpiration force induced at the hot corners is the origin of these secondary vortices.

For $Kn = 0.3$, unlike first order solution which is not much sensitive to the rarefaction increase (Fig. 3b), the second order solution (Fig. 3d) predicts a little change in the center and size of vortices which correspond to the rarefaction effects contribution. In contrast to both NSF approach, the $R13$ solution allows capturing the interplay between thermal stress and transpiration force, which are responsible for the streamlines inversion and disappearance observed at hot corners in the rarefied case. The third term in the $R13$-constitutive equation (9) expresses the thermal stress contribution. This is due to the predominance of non-equilibrium effects, with respect to thermal convection phenomenon, observed in the slightly rarefied case ($Kn = 0.05$). Indeed, the central vortex vanishes and the hot-corners vortices agglomerates to give an inverted circling of gas flow. This is in good agreement with the simplest *windmill* experiment realized by Sone (1991) where the qualitative results are published through a brief communication (Sone 1991) and recorded on a videotape (VHS). In this experiment, a rectangular glass plate of $70 \times 200\ mm$ was set in space with its longer sides in the vertical direction and electrically heated from the lower end of the plate (in back). A windmill was placed in front of the plate to detect the vertical flow induced by the heating process. The system is placed within a glass cylindrical chamber in which the pressure change is well controlled, the experimental protocol is shown in (Fig. 4). At the high-pressure condition, low rarefaction degree, the thermal convection phenomenon provides the windmill motion. This motion becomes increasingly slow with the pressure decrease. At low pressure, $\sim 40 Pa$, the windmill motion reverses by rarefaction effects. Experimentally, he fined as the pressure decreases, downward flow, a flow in the direction of the temperature gradient along the plate, grows and finally overcomes the thermal creep flow. This experiment was performed later with the same qualitative results in various systems with different windmills and plates.

The accuracy of the $R13$ solution, in the rarefied case, is also provided by a wide range of temperature variation, $1.04 \leq T/T_0 \leq 1.24$, unlike the closed one predicted by NSF solutions, especially the second order. The gas flow behavior is, therefore, influenced by both rarefaction and non-isothermal wall effects. To understand more this critical change, observed in the flow streamlines at the rarefied case (Fig.3f), the velocity and heat flux fields are evaluated in the vertical center-line, i.e., $x = L/2$. The following figures (Figs.5 and 6) show the velocity and heat flux components as a function of $y$-coordinate. The rarefaction effects influence strongly the velocity fields, $V_x$ and $V_y$. This higher sensitivity of velocity components is due to the small velocity-magnitude induced only by the longitudinal temperature gradient along the walls. Even at the slightly rarefied case, the classical NSF solutions cannot capture the nonlinear phenomenon contribution, which is more highlighted by solving the regularized 13-moment equations.

This velocity-rarefaction dependence is consistent with the experiment results for different pressure values. The combined effects of rarefaction and non-isothermal wall-driven is proved later by Sone through the improved version of his simple windmill experiment. The results show that the windmill rotation speed is strongly linked to the temperature gradient, i.e., the heating rate $T_0/T_H$, and the rarefaction degree (the pressure value in the chamber).

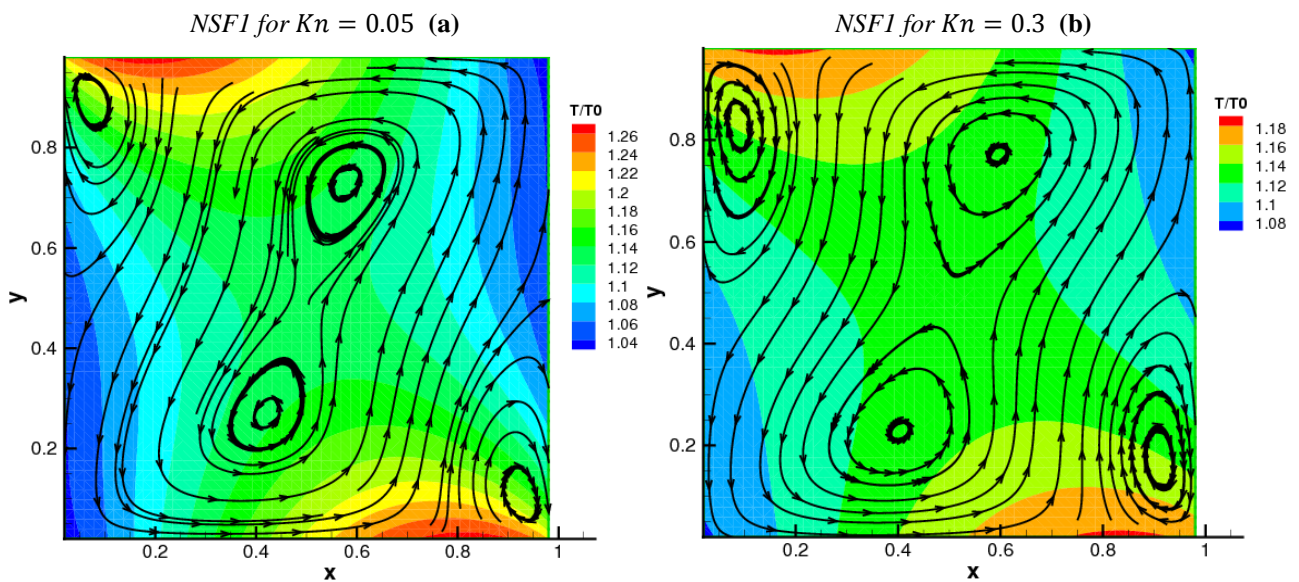

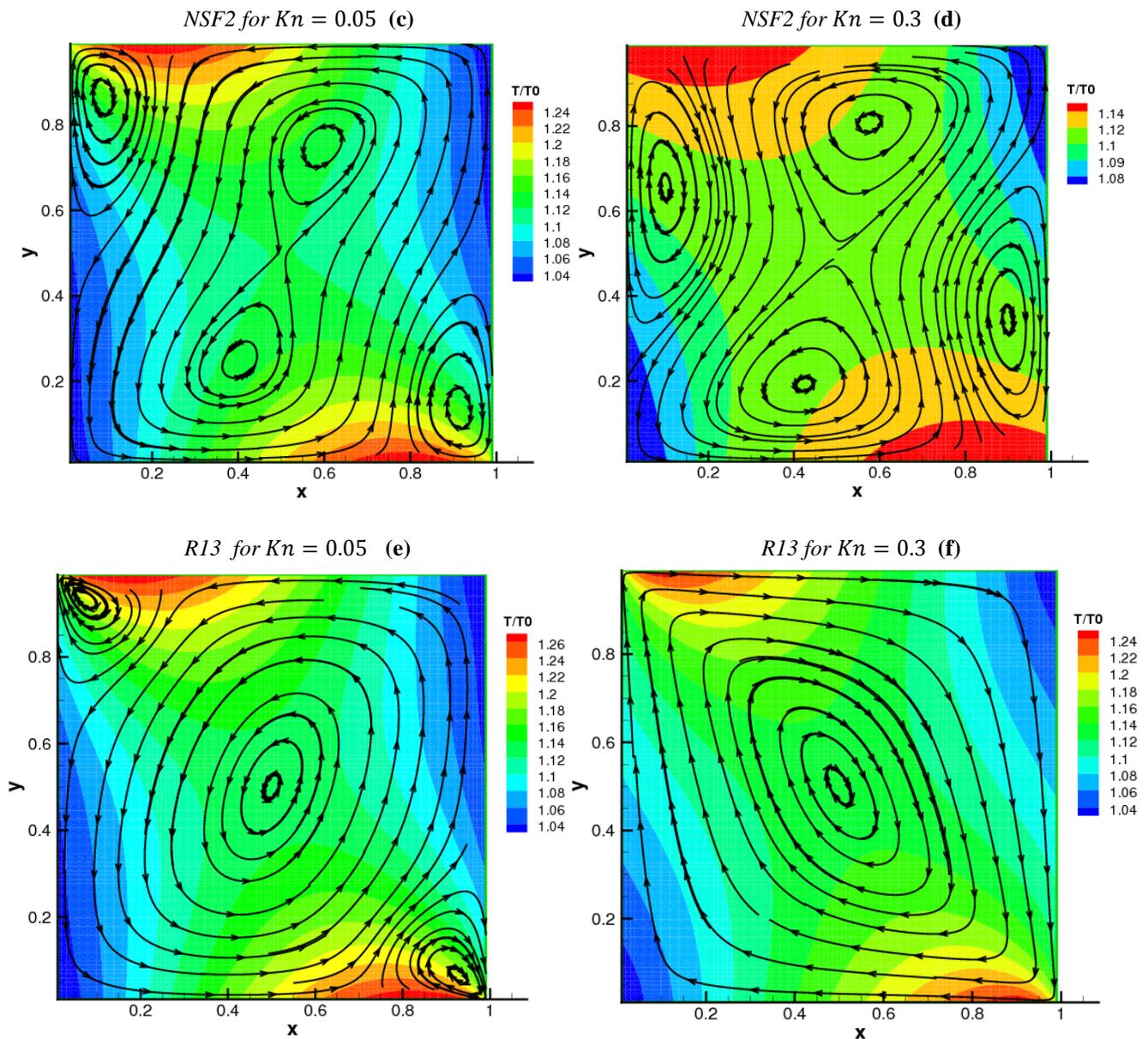

Fig. 3: Streamlines overlaid on the temperature contours for $Kn = 0.05$, and 0.3 using NSF approaches (**a-d**) and R13 approach (**e, f**).

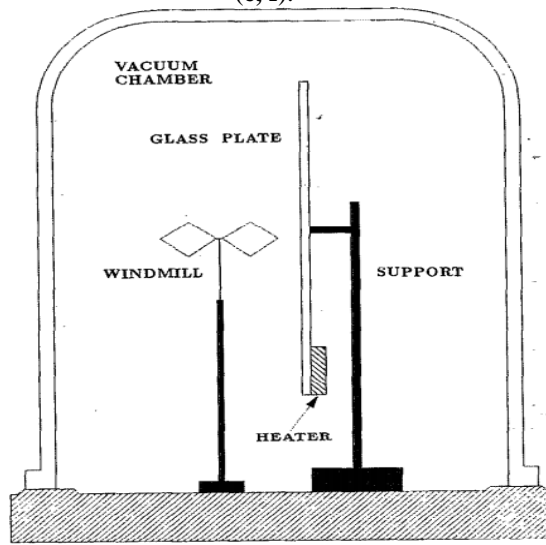

Fig. 4: Windmill experimental apparatus (Sone 1991).

The $R13$ approach provides also a good clarification of the so-called Knudsen-layer which corresponds to the non-equilibrium area induced by the rarefaction effects near the heated walls, located at $y = 0, H$. The rarefaction effects also influence the profiles of heat flux components. At the slightly rarefied case, $Kn = 0.05$, the NSF approaches give rather similar profiles in the bulk region over the range of $0.25 \leq y/L \leq 0.75$. However, the validity of first and second orders BCs of NSF solution decreases with the increase of Knudsen number value before its break down for $Kn = 0.3$. Therefore, the enlargement of the Knudsen-layers in terms of $Kn$ value, induced by the thermal creep flow, is more captured by the R13 solution, unlike the second order NSF2 which predicts only a small correction with respect to the first one NSF1.

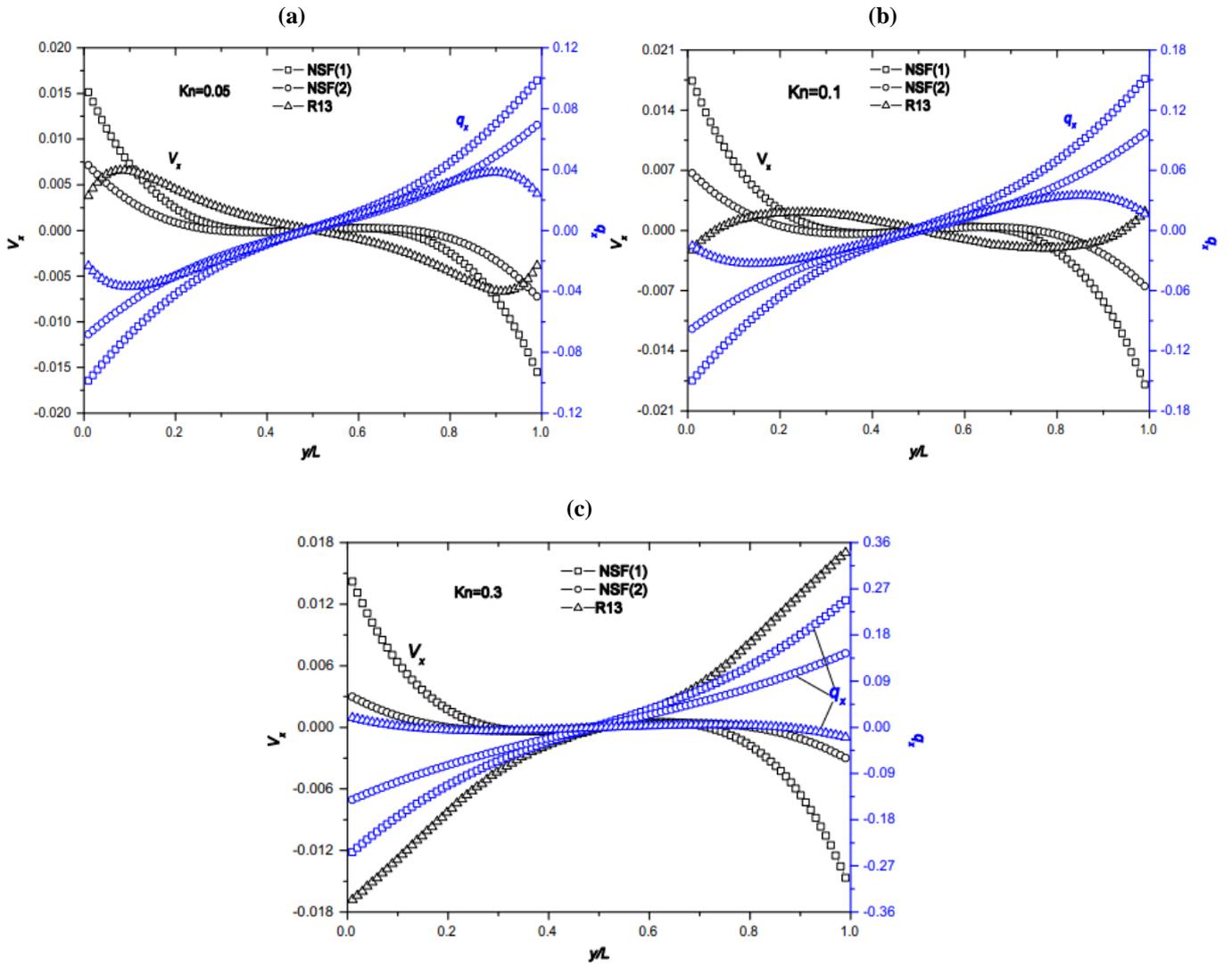

Fig. 5: $x$−velocity and heat flux components profiles at the vertical center-line of the cavity for $Kn = 0.05$ (a), 0.1 (b) and 0.3 (c) using NSF1, NSF2, and R13 approaches, the black and blue symbols denote $x$−velocity and heat flux components respectively.

This non-equilibrium region, which appears in the profiles of longitudinal components as a function of the normal coordinate, $y$, does not take place in the vertical ones. This is illustrated by the $V_y$ and $q_y$ profiles for $Kn = 0.05$ and $0.3$ (Fig. 6). With the second order of velocity slip and temperature jump BCs, the NSF theory predicts an improved correction of $y-$velocity and heat flux components in the rarefied case, see Fig. 6(c), (d). Fig. 6 shows that the three solutions are almost confounded at the slightly rarefied case $Kn = 0.05$. However, for $Kn = 0.3$, the classical NSF theory, especially the first order one, cannot describe correctly the thermally-driven rarefied gas flow behavior. Using the first order of slip and jump BCs, the slip velocity phenomenon is only induced in the proximity of walls, i.e., the Knudsen-layer, by the viscous slip flow. The sufficiency of this order of correction has been proved recently by comparing the R13 results and those obtained by solving the linearized Boltzmann equation for viscous slip and transpiration flow problems (Rana and Struchtrup 2016, Ohwada et al. 1989). However, when the tangential temperature gradient is applied, i.e., $\partial T/\partial x_\tau \neq 0$, contribution of the second order term ($\sim Kn^2$) in the slip BCs must be considered. This explains the accuracy of NSF2 correction in the profile $V_x = f(y)$, with respect to the first one NSF1.

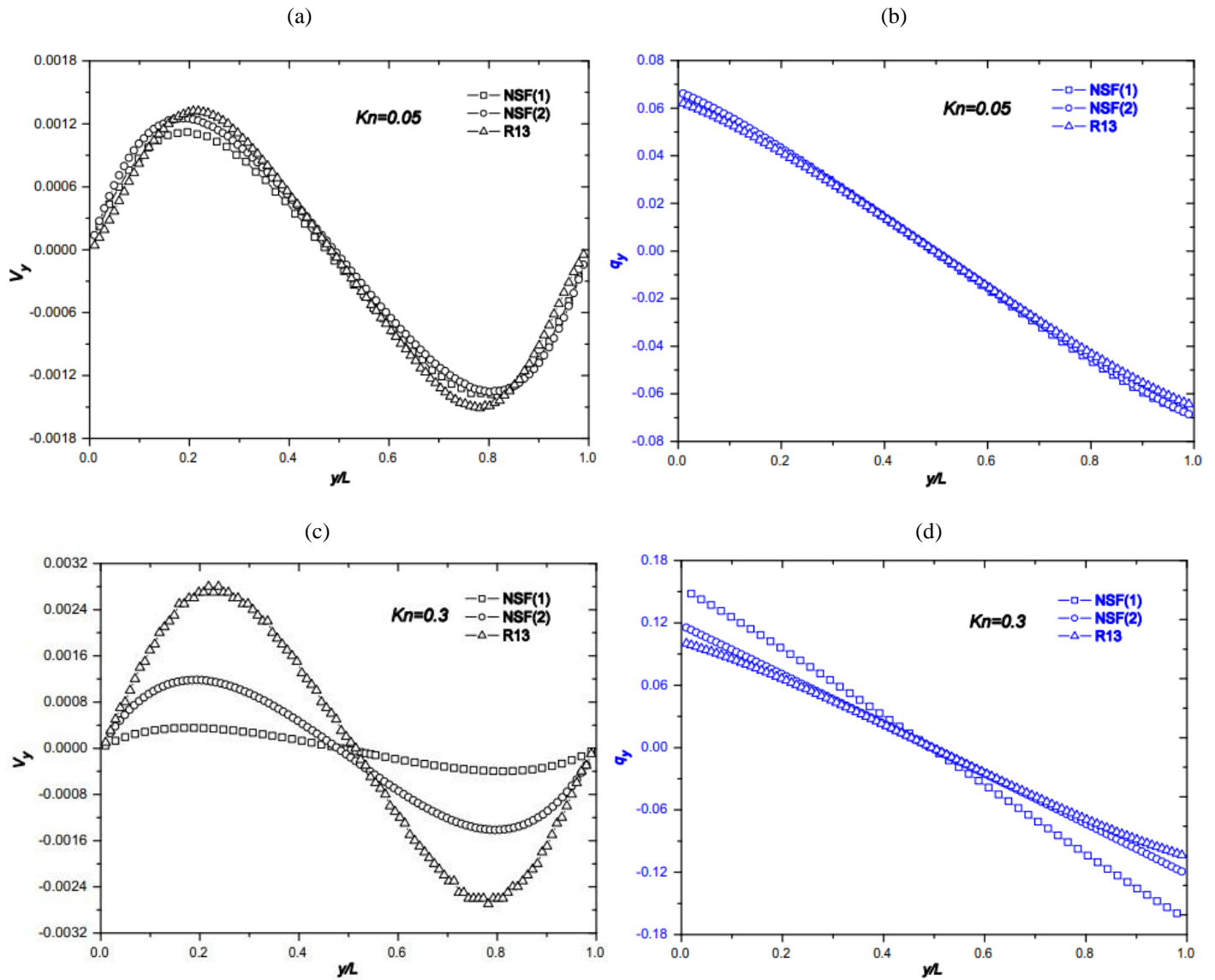

Fig. 6: $y-$ velocity and heat flux components profiles at the vertical center-line for $Kn = 0.05$ and $0.3$ using NSF1, NSF2 and R13 approaches.

To evaluate the combined effects of temperature gradient and rarefaction degree, we show the maximum value of $V_x$ along the bottom wall as a function of Knudsen number for different temperatures, $1.25 \leq T_H/T_0 \leq 2$ (Fig. 7). As a conclusion, the maximum velocity along the wall depends on the gas flow regime defined by the Knudsen number value. In the slip regime, $Kn \leq 0.1$, $V_{x,max}$ decreases with the increase of rarefaction degree. After this limit, transition regime, the gas particles undergo a strong thermal motion at the wall proportional to $Kn$ and $T_H$ values.

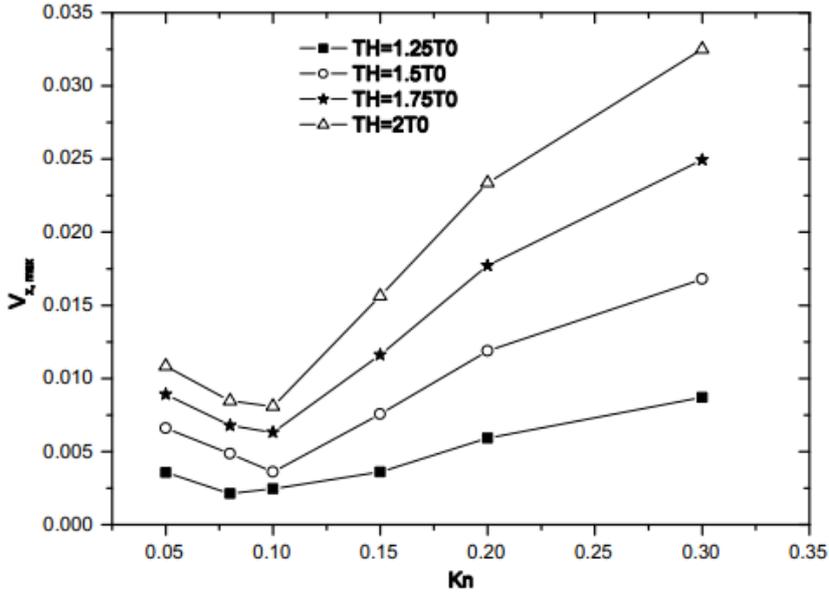

Fig. 7: Maximum $x-$ velocity along the bottom wall for different values of $T_H$

To understand more about the heat transfer process induced by the temperature gradient, the heat lines overlaid on the thermal shear stress contours obtained by NSF2 and R13 approaches have shown in Fig. 8. Comparing the figures, one can conclude that both approaches provide a rather similar distribution of heat flux isolines at the slightly rarefied case ($Kn = 0.05$), while the longitudinal thermal stress magnitude is strongly influenced by the non-linear mechanism of momentum transfer at the boundary. The non-linear phenomena contribution, described with more details from the kinetic point of view, is well considered by the extended continuum theory of R13 by means of the terms $R_{ij}$, $m_{ijk}$, and $\Delta$ unlike the NSF2 solution which gives a small range of change (Ohwada et al. 1989). However, when the Knudsen number becomes larger ($Kn = 0.3$), the rarefaction effects influence the gas flow. The heat flux caused by the gradients of stresses and thermal creep is better described by R13 method (Bobylev 1995, Yariv 2008). The rarefaction effect causes a change in the heat-lines curvature ($Fig.\,9d$), which are not predicted by NSF solution. This is due to the faster motion of particles coming from the hot corners than those coming from the cold ones. The stress quadratic flow, as well as nonlinear thermal stress effects, becomes larger in the flow bulk region (Sone 2007).

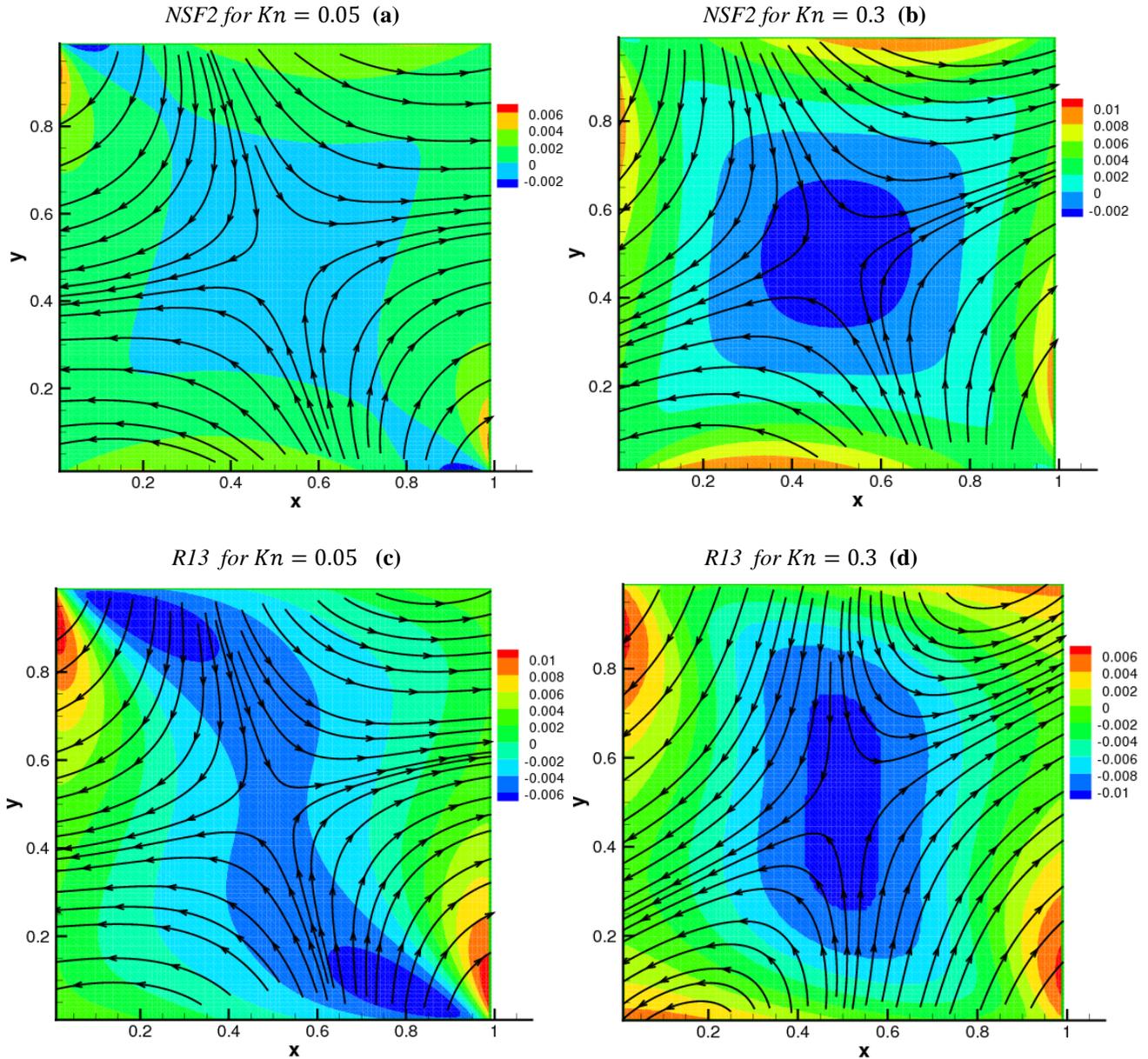
Fig. 8: heat-lines overlaid on the stress $\sigma_{xy}$ contours for $Kn = 0.05$ and $0.3$ using NSF2 (**a**, **b**) and R13 (**c**, **d**) approaches.

The mutual interaction between the longitudinal walls, subject to the temperature gradient, on the flow pattern is evaluated by considering higher cavity ($H = 2L$), i.e., $AR = 2$. Fig. 9 shows the streamlines superimposed on the temperature contours for $Kn = 0.05$ and $0.3$ using both NSF1 and R13 approaches. Comparing these flow fields with the corresponding ones in the square case ($AR = 1$) in Fig. 5, one can conclude that both solutions are sensitive to the cavity aspect ratio and rarefaction effects. For $Kn = 0.05$, the primary vortices observed between the cold corners in the square case, using NSF approach, are detached to form two secondary vortices on the side of lateral walls. But, at the rarefied case, the situation is reversed and the hot-corners vortices begin to agglomerate to form one primary vortex between the hot corners. However, the extended continuum method, R13, predicts rather similar fields but with small edges at the cold corners. For $Kn = 0.3$, two secondary vortices appear, under the vertical flow elongation effect, within the inverted vortex obtained between the hot corners in the square cavity case. These vortices are made to the flow stagnation appeared at $y = 0.5L$ and $1.5L$. On the other hand, the temperature contours show that the R13 approach predicts a larger range of temperature variation, $T/T_0$, in agreement with the rarefaction degree: $[1.04 - 1.24]$ in square case and $[1.02 - 1.22]$ for $AR = 2$. But, with the NSF solution, the temperature range varies from $[1.08 - 1.18]$ to $[1.065 - 1.13]$ in the higher cavity. The sensitivity of thermal behavior with the cavity aspect ratio is, therefore, well captured by the extended continuum theory of R13 than the classical one (NSF).

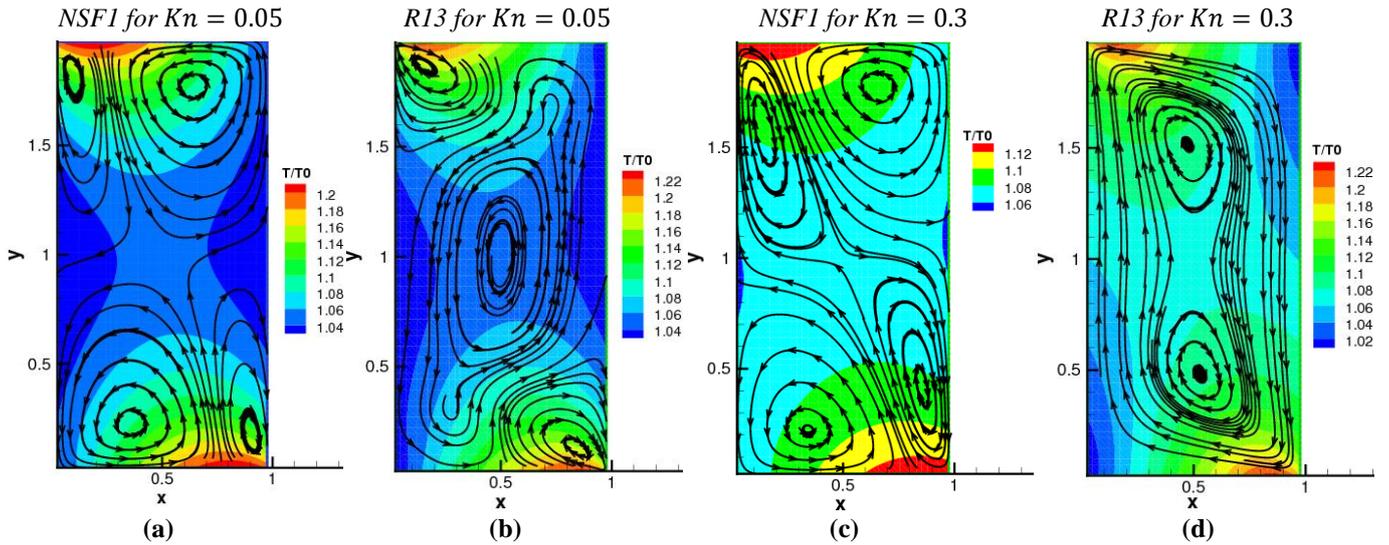

Fig. 9: streamlines overlaid on the temperature contours at $Kn = 0.05$ and $0.3$ for $AR = 2$ using NSF1 and R13 approaches.

**CONCLUDING REMARKS:**

In the present study, the behavior of a rarefied gas flow induced thermally by non-isothermal wall effects within a microcavity is investigated in the slip and early transition regimes. The problem is treated in the frame of classical and extended macroscopic theory by solving numerically, on one hand, the Navier-Stokes and Fourier equations using the first and second orders of velocity slip and temperature jump and the R13 set of equations, on the other. The gas flow sensitivity to the rarefaction degree is clarified using the main macroscopic flow fields. The results prove the break downs of NSF solutions at slip-transition regimes limit. Thus, the extended continuum solution allows the capture of the inverted motion of rarefied gas flow created by the thermal creep flow at the walls. This critical change in the flow streamlines is in a good agreement with the *windmill* experiment results realized by Sone. The non-linear thermal stress and thermal creep co-contribution on the flow fields at the rarefied case have been illustrated by means of the R13 method. In view of different geometries that one can meet in the several industrial applications, two different aspect ratios are evaluated in our study.

**FUNDING STATEMENT:**

We declare that the present work is not object of any funding or supporting body.

**ACKNOWLEDGMENT:**

The authors gratefully acknowledge the valuable contribution of Mr. Vahid Shahabi, member of High Performance Computing (HPC) Laboratory, Department of Mechanical Engineering, Ferdowsi University of Mashhad, Mashhad, Iran, in the discussion of results and editing of the manuscript.